\documentclass[a4paper]{jpconf}
\usepackage{graphicx}
\usepackage{citesort}
             \usepackage{color}
\begin{document}
\title{Fusion excitation function revisited}

\author{Ph Eudes$^1$, Z Basrak$^2$, F S\'ebille$^1$, V de la Mota$^1$, 
G Royer$^1$ and M Zori\'c$^2$}

\address{$^1$ SUBATECH, EMN-IN2P3/CNRS-Universit\'e de Nantes, P.O.Box 20722, F-44\,307 Nantes, France}
\address{$^2$ Rud\hspace*{-0.76ex}{\raise 1.22ex\hbox{\vrule height 0.06ex width 0.28em}}er
   Bo\v{s}kovi\'c Institute, P.O.Box 180, HR-10\,002 Zagreb, Croatia}

\ead{eudes@subatech.in2p3.fr}

\begin{abstract}
We report on a comprehensive systematics of fusion-evaporation
and/or fusion-fission cross sections for a very large variety
of systems over an energy range 4$A$--155$A$ MeV.
Scaled by the reaction cross sections, fusion cross sections
do not show a universal behavior valid for all systems
although a high degree of correlation is present 
when data are ordered by the system mass asymmetry.
For the rather light and close to mass-symmetric systems the
main characteristics of the
complete and incomplete fusion excitation functions can be
precisely determined.
Despite an evident lack of data above 15$A$ MeV for
\textit{all} heavy systems the available data suggests that
geometrical effects could explain the persistence of incomplete
fusion at incident energies as high as 155$A$ MeV.
\end{abstract}

\section{Introduction}
Three regimes of fusion process only qualitatively explain the fusion
excitation function\ \cite{bas80}.
The two low energy regimes have been thoroughly studied in the
past as well as today\ \cite{st-malo}.
On the contrary, the third high-energy regime related to the
extinction of fusion cross section has been studied only
occasionally.
The question \textit{how and why} the fusion process disappears
did not receive a satisfactory answer yet.
The two systematics on fusion excitation function data have
been published so far,
in 1984 by Mogenstern et al.\ \cite{morgenstern84}  and by the 
Indra collaboration in 2006\ \cite{lautesse06} 
both being focused on rather light and symmetric systems. 
In the former study, the disappearance of complete
fusion has been foreseen for $v_L\!<\!0.19c$ and the onset of
incomplete fusion for $0.06c\!<\!v_L$\footnote{$v_L$ denotes
the velocity of the lighter partner in the center-of-mass frame.}. 
For a symmetric system and in terms of incident energy these
limits are respectively equal to 67$A$ MeV and 6,7$A$ MeV.
In the present work we challenge both previously established
limiting values.
In the latter systematics, presented is a fusion excitation
function for 7 light nearly symmetric systems.
It has been found that quasi-fusion becomes vanishingly small 
above 50$A$ MeV.

In this work we have considerably extended the
previously published systematics and an exhaustive 
review of measured fusion cross sections
available today at energies well above the Coulomb barrier
($E_{\rm LAB}\!>\!4A$ MeV) is presented.
Undertaken a scrutiny scan of the published fusion data over
the last forty years ended with 168 fusion cross section
(FCS) values belonging to 57 different systems:
A span in total system mass $A_{\rm tot}\!=\!A_t\!+\!A_p$
($A_t$ and $A_p$ stand for target and projectile mass,
respectively) ranges between 26 and 246, system mass asymmetry
$\mu\!=\!|A_t\!-\!A_p|/(A_t\!+\!A_p)$ from 0 to 0.886,
neutron to proton $N/Z$ ratio from 1 to 1.536 while incident
energy lies between 4$A$ and 155$A$ MeV.

Created high-energy excitation function for the fusion-evaporation
and/or fusion-fission processes should allow 1) to pin down a possible
regularity in the global available data set or in appropriately
selected data subsets, 2) to identify a possible
missing data and to establish a priority list of new measurements
that would have to be performed and 3) to deliver a set of data that
can be used to constrain microscopic transport models used to
describe properties of heavy-ion collisions around the Fermi energy.
In fact, reproducing the overall evolution as well as the main
features (onset, maximum and extinction) of the global fusion
excitation function or the one obtained by sorting data as a
function of a given entrance channel parameter is a real challenge
and a \textit{sine qua non} condition for these models.
Indeed, before pretending to describe in
detail properties of the reaction exit channel such as creation
and properties of particle species with their energy spectra
and isospin these models firstly have to accurately reflect the
global features of reaction mechanisms involved.

\section{The raw data}
\begin{figure}[bth]
\includegraphics[width=23.4pc]{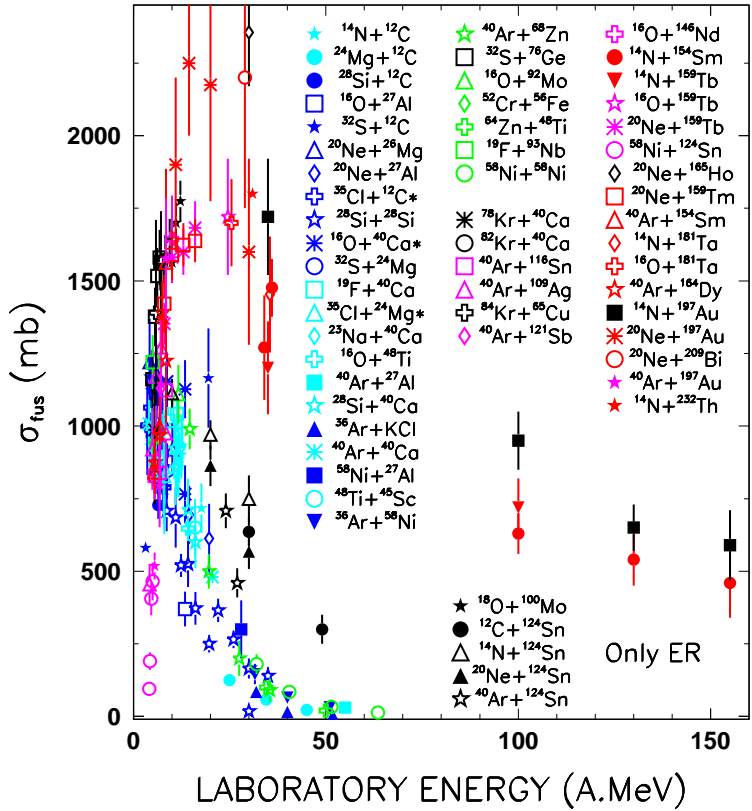}\hspace{1.5pc}%
\begin{minipage}[b]{13.0pc}\caption{\label{stp1}(Color online.)
Raw fusion cross sections plotted as a function of the
laboratory energy per nucleon $E_{\rm LAB}/A$.
The inventoried systems are distinguished among them by a
color code which is used throughout this paper. For convenience, 
all the references are indicated in Figs.\ \protect\ref{stp3}
and\ \protect\ref{stp4}.
See text for details.}
\end{minipage}
\end{figure}

In Fig.\ \protect\ref{stp1} are displayed all collected fusion
data as a function of incident energy in the laboratory
reference frame sorted by the increasing system mass
$A_{\rm tot}$.
The systems are differentiated by a color code (online):
blue and green symbols label the lighter systems
($26\!<\!A_{\rm tot}\!<\!116$) while pink and red symbols
the heavier ones ($146\!<\!A_{\rm tot}\!<\!246$).
Black symbols are used to point out measurements for which
in the original publication is explicitly mentioned a
possible overestimation of the fission component due to an
unresolved contribution of fast-fission\
\cite{guillaume82,ademard11,britt76,songzoni96}.
Black symbols are also used for the five systems labeled by
\textit{Only ER} in Fig.\ \protect\ref{stp1} for which the
evaporation cross sections were measured only, whereas a
more or less significant fission component was expected too\
\cite{kelly97,nifenecker85}.

It is instructive to investigate the systems by sorting them
following both their mass asymmetry and their isospin.
By a careful survey of the list of systems
one infers easily that lighter systems (top of the list) are
rather symmetric both in mass and isospin 
($0\!<\!\mu\!<\!0.5$ and $1\!<\!N/Z\!<\!1.25$) while, on the
contrary, heavier systems (bottom of the list) are very
asymmetric in mass (for most of them $\mu\!>\!0.75$) and, 
as expected, more asymmetric in isospin ($N/Z\!>\!1.3$). 
The ensemble of fusion data splits into the two distinct sets
exactly matching our color code. 
A first set corresponds to Light Symmetric Systems (LSS) for
which FCS diminish with incident energy and disappear 
above 40--50$A$ MeV (see also\ \cite{lautesse06}).
A second set corresponds to Heavy Asymmetric Systems (HAS)
for which FCS increase up to 20$A$ MeV and then decrease.
The HAS branch surprisingly persists at energies as high as
155$A$ MeV. 
At that point one must stress an evident lack of FCS data:
For the heaviest systems in general data are scarce above
15$A$ MeV, particularly for HAS between 30$A$ and 100$A$
MeV, and for systems of medium mass asymmetries in the
entire energy range. 
Let us add that fusion data above 100$A$ MeV are from a single 
experiment \cite{songzoni96}.
  
\begin{figure}[bth]
\includegraphics[width=23.4pc]{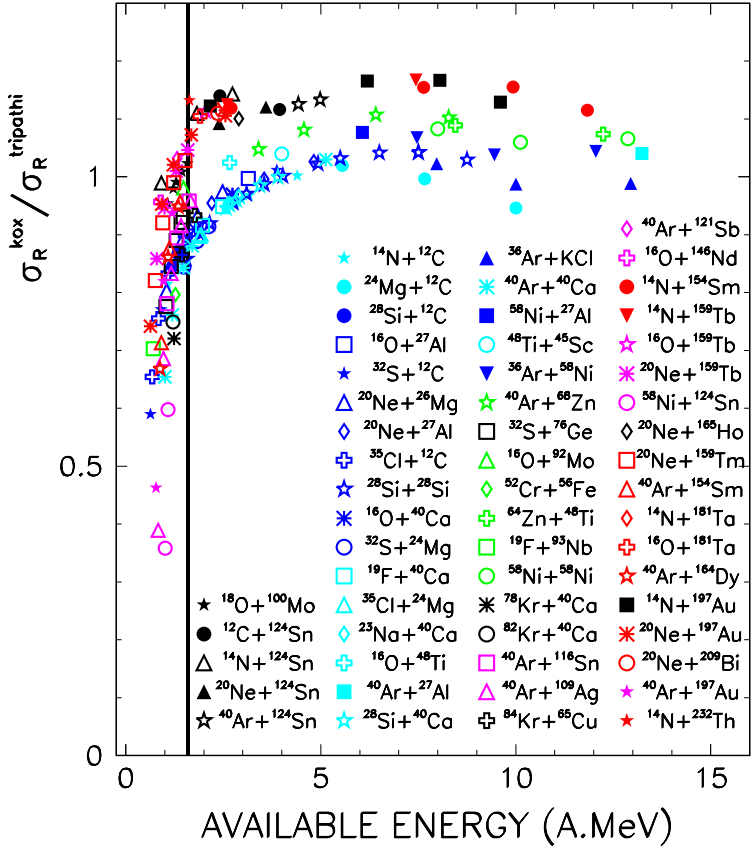}\hspace{1.5pc}%
\begin{minipage}[b]{13.0pc}\caption{\label{stp2}(Color online.)
Kox over Tripathi reaction cross section 
$\sigma_{\rm R}^{Kox}/\sigma_{\rm R}^{Tripathi}$.
The black vertical line indicates the 
$E_{\rm c.m.}/A\!=\,$1.6$A$ MeV limit.
See text for details.}
\end{minipage}
\end{figure}

\section{Normalization of the fusion cross sections}
To reduce the effect of $A_{\rm tot}$ one normalizes FCS by
reaction cross sections $\sigma_{\rm fus}/\sigma_{\rm R}$.
By specifying abscissae in units of available energy
(\textit{i.e.}\ the system center-of-mass energy per nucleon
$E_{\rm c.m.}/A$) one expresses the mass asymmetric systems
on the same footing with those which are mass symmetric.
These coordinate scalings greatly ease the comparison of
various systems with each other.
Several parameterizations are used in calculation of
reaction cross sections\ \cite{tripathi,kox,shen,sihver,shukla,ibrahim}. 
The parameterization of Tripathi is chosen because it is
applicable for any combination of colliding nuclei and valid
over the large energy range from a few $A$ MeV to a few $A$
GeV\ \cite{tripathi}. 
Note that this parameterization depends on the
root-mean-square charge radii of nuclei for which we used
the tabulated values of Ref.\ \cite{angeli04}.
Nevertheless, to check that our results do not depend too
much on the chosen prescription, a comparison between the
Kox\ \cite{kox} and Tripathi formulas has been performed,
Kox formula being modified in order to extend its applicability 
down to a few $A$\,MeV similarly as Shen et al.\ have done\ \cite{shen}. 
The comparison is displayed in Fig.\ \protect\ref{stp2} where is shown the ratio
$\sigma_{\rm R}^{Kox}/\sigma_{\rm R}^{Tripathi}$
of reaction cross sections calculated respectively by Kox and Tripathi formulas.
Again, the two sets previously mentioned (LSS and HAS) are evident.
For HAS, it is clear that this ratio is almost constant, varying
between 1.09 and 1.17 for 
$E_{\rm c.m.}/A\!>\,$1.6$A$ MeV.
Below this limit denoted by the thick vertical line in Fig.\
\protect\ref{stp2}, the differences are growing and reach more than 
60\,\% at the lowest energies.

For LSS, the conclusion is similar except that 
the ratio is closer to 1 (within 10\,$\%$).
Subsequently, we will restrict the domain of validity of our
analysis to $E_{\rm c.m.}/A\!>\,$1.6$A$ MeV
where both formulas are quite compatible.
Thus, choosing Tripathi formula as we have done, implies a slight difference 
in absolute values of normalized cross sections but does not alter the shape 
of the excitation function. 
Anyway, it is likely that none of these parameterizations is
correct at such low energies.
To recall this zone of inconsistency of evaluated total reaction
cross sections $\sigma_{\rm R}$ in all next figures a
yellow band is added for $E_{\rm c.m.}/A\!<\,$1.6$A$ MeV.

\begin{figure}[bth]
\begin{minipage}{18.2pc}
\includegraphics[width=18.2pc]{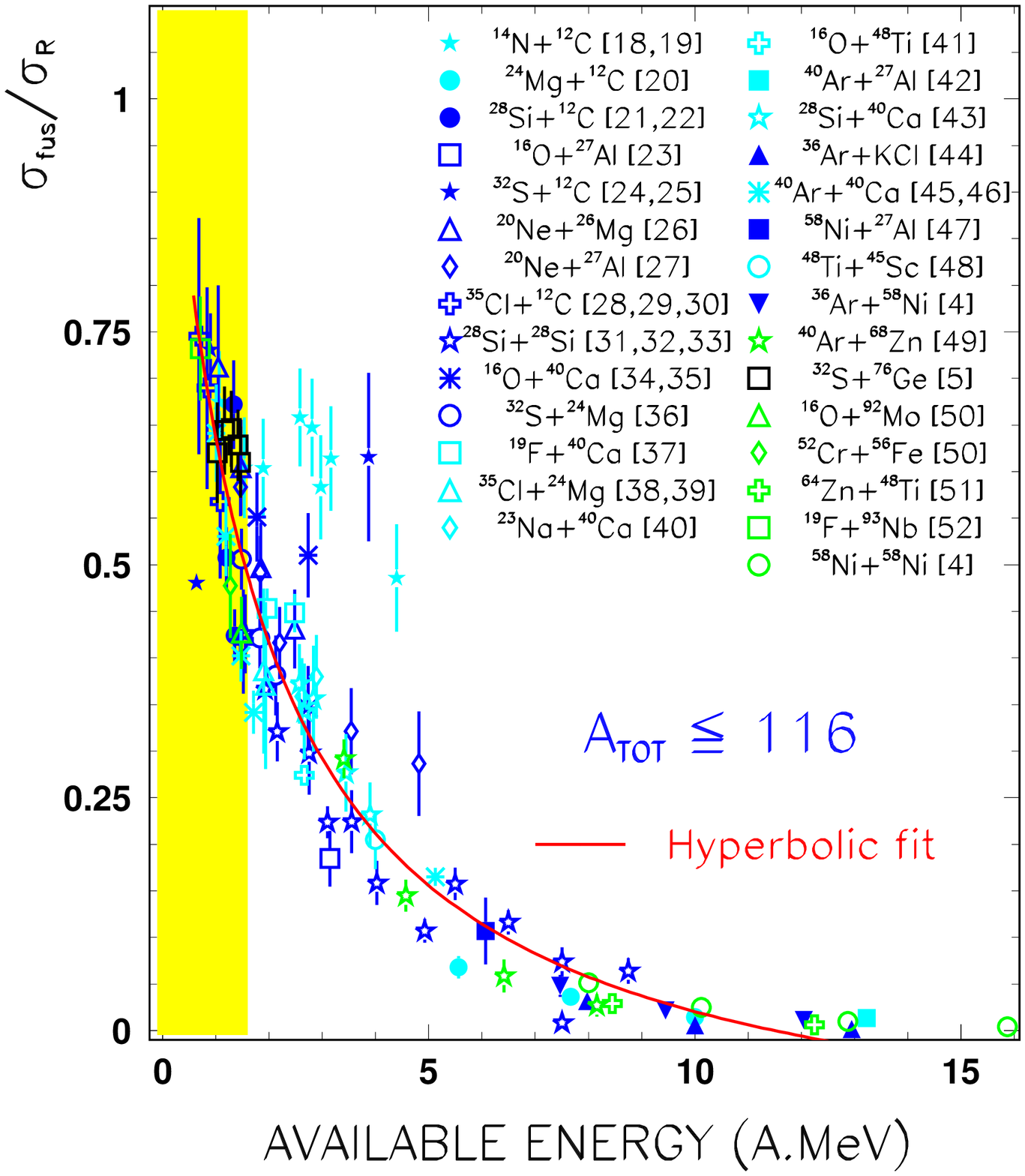}
\caption{(Color online.)
Normalized fusion cross sections for lighter and rather
symmetric systems plotted as a function of $E_{\rm c.m.}/A$. 
See text for details. 
\label{stp3}}
\end{minipage}\hspace{1.5pc}%
\begin{minipage}{18.2pc}
\vspace*{-1.1pc}
\includegraphics[width=18.2pc]{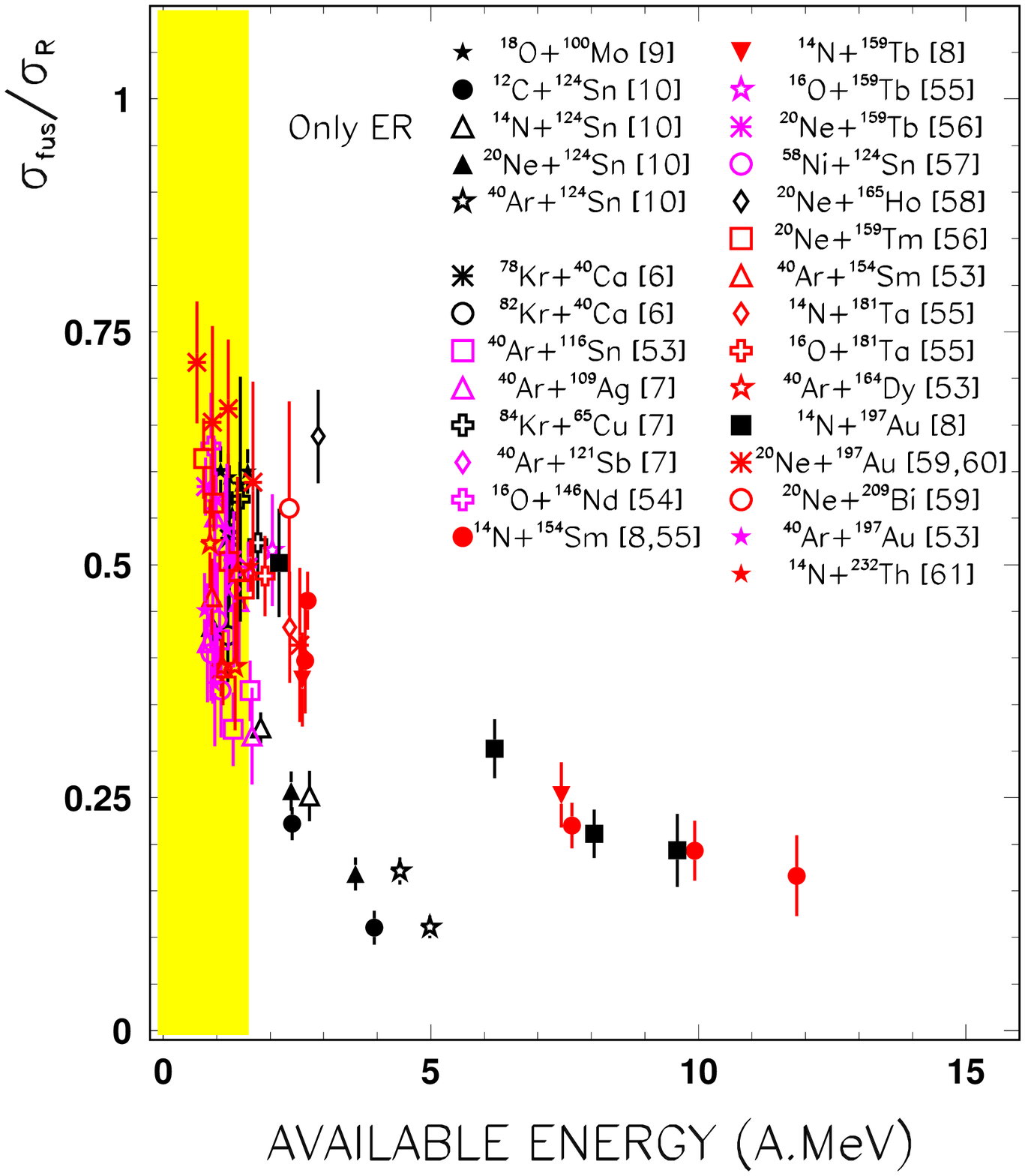}
\caption{(Color online.)
Same as Fig.\ \protect\ref{stp3} for heavier asymmetric systems.
See text for details.
\label{stp4}}
\end{minipage}
\end{figure}

Figures \protect\ref{stp3} and \protect\ref{stp4} show the
LSS and HAS data sets in the scaled coordinates, respectively.
A particularly huge effect is observed for HAS, but clearly,
contrary to what one may expect, FCS cannot be put under a
universal behavior that would be valid for all systems.
In fact, instead of following a single global law, plotted
FCS strongly suggests the existence of two branches: a
lighter symmetric one (Fig.\ \protect\ref{stp3}) and a
heavier asymmetric one (Fig.\ \protect\ref{stp4}).
These figures call for several comments.
When only LSS are considered (Fig.\ \protect\ref{stp3}),
apart from a few points corresponding to the N\,+\,C and
S\,+\,C systems, a very regular correlation is observed. 
The preliminary FCS data on a symmetric and rather heavy
Xe\,+\,Sn system reported at this Conference\ \cite{chbihi12} 
seems to nicely fall in the mentioned
systematics (not included in Fig.\ \protect\ref{stp3}).
The LSS excitation function can be very well reproduced by
a hyperbolic function (see the solid red curve in Fig.\
\protect\ref{stp3}). 

The fit was achieved with the points located outside the
yellow zone.
The fit parameters, however, are quasi-independent on
whether or not the points of the yellow zone are taken into
account.
The result is also independent on whether the Kox or
Tripathi formula was used.
Both ${\sigma}_{\rm fus}$ and ${\sigma}_{\rm R}$ depend on
available energy.
No obvious and simple physical explanation justifying the
observed hyperbolic dependence can be drawn using
\textit{e.g.}\ $1/E_{\rm c.m.}$ dependence as suggested in
the Bass model\ \cite{bass74}.
It is probably due to a non-validity of this model at the
highest energies of the so-called regime III of fusion
process (see \textit{e.g.}\ in Ref.\ \cite{bas80}). 
Further important result is that incomplete fusion reaction
mechanism clearly disappears around $E_{\rm c.m.}$\,=\,12$A$
MeV whatever the LSS system is.
This findings is in agreement with the results of Ref.\
\cite{lautesse06}.
By using the microscopic transport Landau-Vlasov model\
\cite{rem-seb} we have shown years ago that the fusion process abruptly 
disappears around the Fermi energy in head-on collisions:
An elongated composite system gives rise to two nuclei of
the reaction exit channel having kept a strong memory of
the entrance channel suggesting a partial transparency
phenomenon\ \cite{eud99}. 

Understanding the HAS data subset seems to be less
straightforward.
Nevertheless, by neglecting the data points of the yellow
zone and those labeled by \textit{Only ER} the few remaining
data suggest the existence of a second asymmetric branch 
of FCS tending towards a constant value at high incident
energies.
Can one explain the persistence of fusion at such high
energies?

Above 100$A$ MeV a crucial role that the reaction geometry
plays in heavy ion collisions is a well established fact.
In such geometrical picture, fusion can occur for collisions
with an impact parameter smaller than $b_{\rm max}$, a value
corresponding to a complete overlap between the projectile
and the target. Otherwise, for impact parameters greater than 
$b_{\rm max}$, part of the projectile pursues its trajectory at 
small angle with a velocity close to the projectile one.
Within this approximation and with a reaction in direct kinematics
the fusion cross section is expressed as following
\begin{equation}
{\bf \sigma}_{\rm fus} = \pi~b_{\rm max}^2 = \pi~(R_t - R_p)^2 
= \pi~r_0^2~(A_t^{1/3} - A_p^{1/3}),
\end{equation}
\noindent
where $R_p$ ($R_t$) is projectile (target) radius.
Assuming the \textbf{simple} geometrical formula for the reaction cross
section 
\begin{equation}
{\sigma}_{\rm R} = \pi~r_0^2~(A_t^{1/3} + A_p^{1/3}),
\end{equation}
\noindent
one gets
\begin{equation}
\left[\frac{{\sigma}_{\rm fus}}{{\sigma}_{\rm R}}\right]_{lim}
= \left[1-\frac{2}{1+(\frac{A_t}{A_p})^{1/3}}\right]^2
= \left[1-\frac{2}{1+(\frac{1+\mu}{1-\mu})^{1/3}}\right]^2
\label{gcs}
\end{equation}
\noindent 
Eq.\ (\ref{gcs}) indicates that at high energies normalized 
FCS tend towards a value which only depends on the mass
asymmetry $\mu$. 
For the $^{14}$N\,+\,$^{154}$Sm and $^{14}$N\,+\,$^{197}$Au
Eq.\ (\ref{gcs}) gives 14.4\,\% and 17.0\,\%, respectively
in the very good agreement with experimental data.
This reaction picture is confirmed by our Landau-Vlasov
simulations\ \cite{eudes13}: For impact parameters of complete overlap 
collisions lead to the formation of a massive incomplete
fusion nucleus accompanied by pre-equilibrium emission of
light particles in the forward direction, an emission
mostly originating from the projectile.
Obviously, for a more and more symmetric system this limit
tends to zero in accordance with what is observed
experimentally.
The mass asymmetry $\mu$ of the system seems to be the key
parameter in explaining the above transition, at least in
the high energy part.
The data on systems of medium $\mu$ are needed to confirm
or deny the above assumption.
\begin{figure}[bth]
\includegraphics[width=23.4pc]{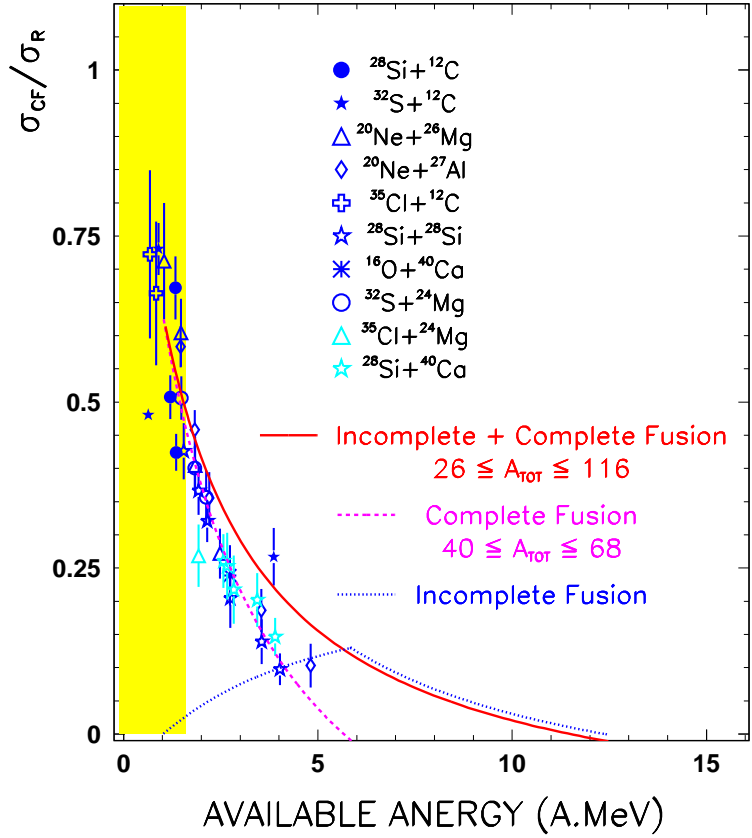}\hspace{1.5pc}%
\begin{minipage}[b]{13.0pc}\caption{\label{stp5}(Color online.)
Normalized complete fusion cross sections for lighter
symmetric systems plotted as a function of the available
energy $E_{\rm c.m.}/A$.
See text for details.}
\end{minipage}
\end{figure}

\section{Complete fusion excitation function}
Among the experiments listed in our systematics only ten
have explicitly been designed to measure both complete and
incomplete fusion components\
\cite{vineyard93,kolata85,lehr84,morgenstern83,pirrone97,vineyard90,beck89,hinnefeld87,cavallaro98,vineyard92}.
This allows establishing a complete fusion excitation
function incorporating 29 FCS data points relative to 10
systems spanning a total mass range from 40 to 68.
Data are reported in Fig.\ \protect\ref{stp5}.
Again, a very regular correlation is observed with FCS
decreasing rapidly as a function of $E_{\rm c.m.}/A$.
As before, it can be fitted by the same kind of hyperbolic law
(see pink solid curve in Fig.\ \protect\ref{stp5}) which
crosses abscissa around 6$A$ MeV.
For the complete fusion reaction mechanism, center-of-mass
energy and excitation energy are equivalent.
Thus, the energy limit of 6$A$ MeV corresponds to the maximal
excitation energy which can be deposited into a light
compound nucleus. 
The above conclusion rises several questions.
What happens for heavier systems? What for asymmetric systems?
Does one obtain the same behavior and the same limit for
complete fusion independently of $A_{\rm tot}$?
Again, further measurements would be welcome.

Finally, in Fig.\ \protect\ref{stp5} are superimposed the two
hyperbolic fits obtained for complete fusion (CF) and for the
sum of CF and incomplete fusion (IF) components and which are
denoted by pink and red curves, respectively as well as their
difference which is colored blue.
These functions provide a measure of the average contribution
of CF and IF fusion components and of their evolution with
energy.
From the results of Fig.\ \protect\ref{stp5} one can, in
particular, infer
\begin{itemize}
\item
 The threshold of IF is around $E_{\rm c.m.}/A$ equal to 1$A$ MeV. 
\item
 By increasing $E_{\rm c.m.}/A$ CF decrease while IF increase.
  At 4$A$ MeV each of these
  two components $\sigma_{\rm IF}$ and $\sigma_{\rm CF}$
  contribute about the same value of $\approx\,0.1\times\sigma_{\rm R}$
  to the total reaction cross section.
\item
 IF component increases up to 6$A$ MeV where it reaches a
  maximum while at the same time the CF process vanishes.
  $\sigma_{\rm IF}$ reaches 15\% of $\sigma_{\rm R}$.
\item
 After this maximum IF decreases and disappears around 12$A$ MeV.
\end{itemize}

To our knowledge, it is the first time that such accurate
limits concerning both CF and IF fusion components are
deducted from such a large body of data.
It allows to draw a simple picture for the evolution of
these reaction mechanisms.

\section{Conclusions}
In summary, we studied the energy dependence of fusion
cross section aiming at understanding the evolution of the
underlying reaction mechanisms as a function of entrance
channel parameters and the causes of its (non-)disappearance. 
The heart of this work is a systematics and as wide as
possible overview of measured fusion-evaporation and/or
fusion-fission cross-sections including summed up complete
and incomplete fusion contributions.  
Normalized by reaction cross sections and plotted as a
function of the available energy per nucleon, the fusion
cross sections show a rather universal behavior leading to a
disappearance of the fusion around 12$A$ MeV in the case of
light symmetric systems.
In the case of heavy
asymmetric systems incomplete fusion seems to persist and tends
towards a constant value depending only on the mass asymmetry 
of the system.
In addition, dissociating available complete and incomplete
fusion cross sections allows to provide fairly accurate
energy limits characterizing these two mechanisms.
Both the observed trends of fusion excitation function and the above 
limits place important constraints on the basic ingredients of microscopic
transport models.

\end{document}